\documentclass[12pt]{article}
\usepackage[a4paper, total={7in, 10in}]{geometry}
\usepackage[parfill]{parskip}
\usepackage{physics, tensor, float, subcaption}
\usepackage{graphicx}
\graphicspath{ {Plots/} }
\usepackage{jhep-mod}
\usepackage{bm}
\usepackage{soul}
\usepackage{amssymb,amsmath,amsthm}
\usepackage{mathrsfs}
\usepackage[utf8]{inputenc}
\usepackage{enumerate}
\usepackage{bigints}
\usepackage{xcolor}
\usepackage{appendix}
\usepackage{graphicx}
\usepackage{float}
\usepackage{tikz}
\usepackage{setspace}
\usepackage{cancel}
\usepackage{subcaption}
\definecolor{purple}{rgb}{1,0,1}
\definecolor{lime}{HTML}{A6CE39} 

\definecolor{lime}{HTML}{A6CE39}
\newcommand{\orcidicon}{%
	\begin{tikzpicture}
	\draw[lime, fill=lime] (0,0) 
		circle [radius=0.16] 
		node[white] {{\fontfamily{qag}\selectfont \tiny ID}};
	\draw[white, fill=white] (-0.0625,0.095) 
		circle [radius=0.007];
	\end{tikzpicture}
	\hspace{-5mm}
}
\newcommand\orcidThomas{{\href{https://orcid.org/0000-0002-0314-4136}{\orcidicon}}}
\newcommand\orcidAlex{{\href{https://orcid.org/0000-0002-1763-3563}{\orcidicon}}}
\newcommand\orcidFrancisco{{\href{https://orcid.org/0000-0002-9388-8373}{\orcidicon}}}
\newcommand\orcidMatt{{\href{https://orcid.org/0000-0003-1088-6485}{\orcidicon}}}

\newcommand{\e}{\mathrm{e}}
\newcommand{\g}{1-\frac{2m \, \e^{-a/R}}{R}}
\newcommand{\gp}{1-\frac{2m_+ \e^{-a_+/R}}{R}}
\newcommand{\gm}{1-\frac{2m_- \e^{-a_-/R}}{R}}
\newcommand{\gpm}{1-\frac{2m_\pm \e^{-a_\pm/R}}{R}}
\renewcommand{\P}{\mathcal{P}}
\begin{document}
\title{\vspace{-25pt}\huge{
Thin--shell traversable wormhole crafted from a regular black hole with asymptotically Minkowski core
}}
\author{
\Large
Thomas Berry\!\orcidThomas$^1$, Francisco S. N. Lobo\!\orcidFrancisco$^2$,\\
Alex Simpson\!\orcidAlex$^1$, {\sf  and} Matt Visser\!\orcidMatt$^1$}
\affiliation{$^1$School of Mathematics and Statistics, Victoria University of Wellington, \\
\null\qquad PO Box 600, Wellington 6140, New Zealand.}
\affiliation{$^2$Instituto de Astrof\'isica e Ci\^encias do Espa\c{c}o, \\
\null\qquad Faculdade de Ci\^encias da Universidade de Lisboa, Edif\'icio C8, Campo Grande, \\
\null\qquad P--1749--016, Lisbon, Portugal.}
\emailAdd{thomas.berry@sms.vuw.ac.nz}
\emailAdd{fslobo@fc.ul.pt}
\emailAdd{alex.simpson@sms.vuw.ac.nz}
\emailAdd{matt.visser@sms.vuw.ac.nz}
\abstract{
\vspace{1em}

Recently, a novel model for a regular black hole was advocated which possesses an asymptotically Minkowski core implemented via an exponential suppression (in the core region) of the Misner--Sharp quasi--local mass.
Using this regular black hole as a template, we shall construct a spherically symmetric thin-shell traversable wormhole using the ``cut-and-paste'' technique, thereby constructing yet another black hole mimicker.
The surface stress-energy at the wormhole throat is calculated, and the stability of the wormhole is analyzed.
An important result is that, (as compared to their Schwarzschild thin-shell counterparts),  increasing the exponential suppression of the Misner--Sharp quasi-local mass by increasing the suppression parameter $a$, also considerably increases  the stability regions for these thin-shell wormholes, and furthermore minimizes the amount of energy condition violating exotic matter required to keep the wormhole throat open.

\bigskip
\noindent
{\sc Date:} 17 August 2020;  22 August 2002; \LaTeX-ed \today


\enlargethispage{30pt}
\bigskip
\noindent{\sc PhySH:} 
Gravitation
}

\maketitle
\def\tr{{\mathrm{tr}}}
\def\diag{{\mathrm{diag}}}
\def\cof{{\mathrm{cof}}}
\def\pdet{{\mathrm{pdet}}}
\parindent0pt
\parskip7pt
\section{Introduction: Aims and goals}

Recent advances in observational and gravitational wave astronomy projects such as the Event Horizon Telescope~\cite{Akiyama:2019cqa,Akiyama:2019brx, Akiyama:2019sww,Akiyama:2019bqs,Akiyama:2019fyp,Akiyama:2019eap} and the 
LIGO/Virgo Collaboration~\cite{ligo-detection-papers, grav-wave-observations-wiki} (and LISA~\cite{Barausse:2020rsu} in the future) have opened up the exciting possibility of testing gravity in extreme astrophysical regimes. These projects have explored the possibility of observationally distinguishing between the near-horizon physics of classical black holes and possible astrophysical mimickers~\cite{Carballo-Rubio:2018jzw, Carballo-Rubio:2018pmi, Carballo-Rubio:2019fnb, Carballo-Rubio:2019nel}.
In fact, in the context of a binary coalescence, it has been argued that great care should be taken, as it is commonly believed that the ringdown signal provides a conclusive proof for the formation of an event horizon after the merger~\cite{Cardoso:2016rao}. This is based on the standard assumption that the ringdown waveform at intermediate times is dominated by the quasinormal modes of the final object. 

However, it has been shown that very compact objects with a light ring will also display a similar ringdown stage, even when their quasinormal-mode spectrum is completely different from that of a black hole. This analysis proves that the ringdown waveforms indicate the presence of light rings, rather than of horizons, and that only precision observations of the late-time ringdown signal, where the differences in the quasinormal-mode spectrum eventually show up, can definitively be used to rule out exotic alternatives to black holes, and to test quantum effects at the horizon scale~\cite{Cardoso:2016rao,Cardoso:2019rvt}.

This observation motivates continued research in exploring black hole mimickers as possible alternatives to black holes, where the properties in the near-horizon strong gravity region may be different for these two types of objects. However, it is also possible that at infinity one may discriminate black holes from their mimickers~\cite{Lemos:2008cv}.
In fact, a plethora of black hole mimickers have been explored in the literature, such as the Dymnikova models~\cite{Dymnikova:1999cz,Dymnikova:2000zi,Dymnikova:2003vt, Dymnikova:2004qg,Dymnikova:2010zz}, the Mazur--Mottola gravastars and generalizations~\cite{Mazur:2001fv,Mazur:2004fk,Visser:2003ge,Cattoen:2005he, Lobo:2005uf,Chirenti:2007mk,Lobo:2006xt,Harko:2009gc,Lobo:2012dp, MartinMoruno:2011rm,Lobo:2015lbc,Chirenti:2016hzd} and, in particular, thin-shell wormholes~\cite{Visser:1989kg,Visser:1989kh,Poisson:1995sv,Eiroa:2003wp, Lobo:2003xd,Lobo:2015xhk,Lemos:2003jb,Lobo:2004rp, Lobo:2004id,Lemos:2004vs,Sushkov:2005kj,Lobo:2005us,Lobo:2005zu, Lobo:2005yv,Eiroa:2004at,Eiroa:2005pc,Thibeault:2005ha,Rahaman:2006vg, Eiroa:2007qz,Lemos:2008aj,Eiroa:2008hv, 
Eiroa:2009hm,Eiroa:2008ky,Eiroa:2009nn,Mazharimousavi:2010bm,Dias:2010uh, Yue:2011cq,DeBenedictis:2008qm,Ishak:2001az,Wang:2020,Garcia:2011aa,Lobo:2020kxn, Simpson:2019cer}.
Indeed, the latter cases are based on systematic applications of the thin-shell formalism (or Darmois--Israel formalism)~\cite{Sen,Lanczos,Darmois, 
Synge, Lichnerowicz,Israel:1966rt,Musgrave:1995ka}, which we shall concentrate on throughout this work.

Herein, we shall use a novel regular black hole spacetime as a template to construct a traversable wormhole~\cite{Morris:1988cz,Morris:1988tu,Visser:1995cc,Visser:1989ef, Cramer:1994qj,Hochberg:1997wp,Visser:1997yn,Dadhich:2001fu,Boonserm:2018orb,Lobo:2017oab} via the well-known ``cut--and--paste'' technique~\cite{Visser:1989kg,Visser:1989kh}, and then analyse the surface stress-energy at the wormhole throat using the Darmois--Israel formalism~\cite{Darmois,Israel:1966rt} mentioned above. The model spacetime which forms the basis of this construction is a regular black hole with an asymptotically Minkowski core, as discussed in~\cite{Simpson:2019mud}.
This is an example of a metric with an exponential mass suppression, and is described extensively below.

\section{Thin--shell wormhole framework}

\subsection{Background} 

After the renaissance of wormhole physics in the late 1980s~\cite{Morris:1988cz, Morris:1988tu}, there was very rapid progress of investigations into thin-shell wormholes. See for instance references~\cite{Visser:1989kh, Visser:1989kg} and~\cite{Visser:1995cc}.  A relatively recent general analysis and summary can be found in reference~\cite{Garcia:2011aa}. 
A very recent brief and cogent literature survey can be found in~\cite{Lobo:2020kxn}.

The central idea behind thin-shell wormholes is to take two bulk spacetimes, excise two regions with isometric boundaries, and identify the boundaries~\cite{Visser:1989kh, Visser:1989kg}. This is effectively a modification of the abstract mathematical notion of the ``connected sum'' of manifolds, wherein one uses metrical information, not just topological information. Key ingredients of the analysis are the two bulk metrics, the (isometric) induced metrics (intrinsic 3-metrics) on the boundaries (the first fundamental forms), and the extrinsic curvatures of these boundaries in the two bulk spacetimes (the second fundamental forms). On the boundary itself there is a delta-function distribution of stress-energy that is related to the discontinuity in the extrinsic curvatures~\cite{Visser:1989kh, Visser:1989kg} in a very precise and specific manner~\cite{Visser:1995cc}. 

We shall now apply this very general and flexible formalism in the specific case of spherical symmetry, choosing the bulk spacetimes to be specific regular black holes with exponential mass suppression, leading to asymptotically Minkowski cores~\cite{Simpson:2019mud}.

\subsection{Construction}

We start with the spacetime of a regular black hole with an asymptotically Minkowski core given by~\cite{Simpson:2019mud}
\begin{equation}
\dd s^2 = -\left(1-\frac{2m\,\e^{-a/r}}{r}\right)\dd{t}^2 + \left( 1-\frac{2m\,\e^{-a/r}}{r}\right)^{-1}\dd{r}^2 + r^2\left(\dd{\theta}^2 + \sin^2\theta \dd{\phi}^2\right).
\label{metricworm}
\end{equation}
A rather different (extremal) version of this model spacetime, based on nonlinear electrodynamics, has been previously discussed by Culetu~\cite{Culetu:2013}, with follow-up on some
aspects of the non-extremal case in references~\cite{Culetu:2014, Culetu:2015a, Culetu:2015b}. See also~\cite{Junior:2015, Rodrigues:2015}.

This spacetime possesses horizons located at
\begin{equation}
r_H = 2m\;\e^{W\left(-\frac{a}{2m}\right)} = {a\over |W\left(-\frac{a}{2m}\right)|},
\label{horizon}
\end{equation}
where $W(x)$ is the real-valued Lambert $W$ function, which is negative for those negative arguments where it is defined.
Equation \eqref{horizon} implies that an outer horizon and an inner horizon exist, which are obtained by either taking the $W_0$ or the $W_{-1}$ branch of the Lambert $W$ function, respectively.
Note that in order for horizons to be present equation \eqref{horizon} forces the parameter $a$ to lie in the interval $a \in \left(0, {2m}/{\e}\right]$, and in particular $a \leq 2m/\e$~\cite{Simpson:2019mud}.

More specifically, we have
\begin{equation}
r_{H^-} = 2m\; \e^{W_{-1}\left(-\frac{a}{2m}\right)}\,, \qquad r_{H^+} = 2m\; \e^{W_{0}\left(-\frac{a}{2m}\right)}.
\end{equation}
For $a < 2m/\e$ one has $r_{H^+}>a>r_{H^-}$. 
For the specific case of $a=2m/\e$, one has $W\left(-{a}/{2m}\right)\to 
W\left(-1/\e\right)= -1$. Then
the two horizons merge at $r_{H^\pm }= a$ and the regular black hole is extremal.
If $a>2m/\e$, the horizon locations are undefined and we are dealing with a horizonless compact object.

For the purposes of thin--shell construction, if horizons are present, then we shall perform spacetime surgery \emph{outside} the outer horizons, where we have good control over the physics, and hence we shall have a thin-shell located at some $r>r_{H^+}>a$.
If horizons are not present, $a>2m/\e$, then we could in principle perform spacetime surgery at any nonzero value of $r$.

In the following, we will consider two copies of the regular black hole spacetime given by the line element \eqref{metricworm}, and subsequently analyse the manifold formed by surgically removing the regions 
$r \in (0,R(\tau))$, with the surface $R(\tau)$ lying \emph{outside} both \emph{outer} horizons (if present) of each spacetime, and ``gluing'' them together along this new boundary.

\subsection{Energy conditions in the bulk} 

The bulk spacetime has the following stress-energy tensor profile:
\begin{eqnarray}
\rho &=& -p_r = \frac{ma \, \e^{-a/r}}{4\pi r^4}, 	
\label{rhopr}
\\
p_t &=& - \frac{ma(a-2r) \e^{-a/r}}{8\pi r^5},
\end{eqnarray}
where $\rho$ is the energy density, $p_r$ and $p_t$ are the pressures in the radial and tangential directions, respectively.

It is interesting to analyse the pointwise energy conditions~\cite{Visser:1995cc, Visser:1997tq, Visser:1997qk, Visser:1999de, Martin-Moruno:2013wfa, Martin-Moruno:2013sfa, Martin-Moruno:2017exc}. Specifically,
in order to satisfy the null energy condition (NEC) we require \( \rho + p_r \geq 0 \) and \( \rho + p_t \geq 0 \). 
Indeed, we have \( \rho + p_r = 0 \) globally, however 
\begin{equation}
\rho + p_t = \frac{r}{2}\rho' = \frac{ma \, \e^{-a/r}}{8\pi r^5}(a-4r),
\end{equation}
and so the NEC is only satisfied in the region \( r \leq a/4 \). 
In view of the fact that the outer horizon (if it exists) is located at $r_{H^+} = 2me^{W_{0}\left(-\frac{a}{2m}\right)}>a$, corresponding to possible locations $r_{H^+}\in\left(a, +\infty\right)$, and we `chop' the spacetime outside any horizons that are present, we may conclude that the transverse NEC is manifestly violated in the bulk regions of the constructed spacetime.\footnote{If horizons are not present, then one might be able to satisfy the NEC for small enough $r$.}
\enlargethispage{20pt}

One of the constraints of the strong energy condition (SEC) is that \( \rho + p_r + 2p_t \geq 0 \).
That is, we require 
\begin{equation}
\rho + p_r + 2p_t = \frac{ma(2r-a) \e^{-a/r}}{4\pi r^5} \geq 0.
\end{equation}
It can be clearly seen that this is only satisfied in the region \( r \geq a/2 \) and so, (regardless of whether or not horizons are present), there is no region in which both the {NEC} and the {SEC} are simultaneously satisfied. However, in the
presence of horizons, this aspect of the SEC will be globally satisfied in the bulk regions.\footnote{If horizons are not present, then one might be able to violate the SEC for small enough $r$.}

\subsection{Four-velocity, unit normal, and extrinsic curvature of the throat}

We now allow the boundary surface \( \Sigma \) to be dynamic. 
For tractability, we consider dynamic perturbations to the radial location of the wormhole throat \emph{only}. 
It follows that the intrinsic metric on \( \Sigma \) is given by:
\begin{equation}
\dd s_\Sigma^2 = -\dd \tau^2 + R(\tau)^2\; (\dd \theta^2 + \sin^2\theta \dd \phi^2),
\end{equation}
with coordinate chart \( x^\mu(\tau,\theta,\phi) = (t(\tau), R(\tau), \theta, \phi) \), where $\tau$ is the proper time of an observer comoving with $\Sigma$.
The implied form for the four-velocity of an observer (or a  piece of stress--energy) located on the junction surface is thus:
\begin{equation}
U^\mu_{\pm} = \left( \dv{t}{\tau}, \, \dv{R}{\tau}, \, 0, \, 0 \right),
\end{equation}
and takes the following explicit form
\begin{equation}
U^\mu_\pm = \left( \frac{\sqrt{\gpm + \dot{R}^2}}{\gpm}, \, \dot{R}, \, 0, \, 0 \right).
\end{equation}

The hyper-surface $\Sigma$ is defined by the function $f(x^\mu(\xi^i))=r-R(\tau)=0$, and so the unit normals to this surface are defined by
\begin{equation}
n_\mu = \pm \abs{ g^{\alpha\beta}\frac{\partial f}{\partial x^{\alpha}} \frac{\partial f}{\partial x^{\beta}} }^{-\frac{1}{2}} \pdv{f}{x^\mu}.
\label{normalSigma}
\end{equation}
A trivial but quite lengthy calculation yields the following unit normal vector to~$\Sigma$:
\begin{equation}
n^{\mu} = \pm\left( \frac{\dot{R}}{\gpm}, \, \sqrt{\gpm +\dot{R}^2}, \, 0, \, 0 \right).
\end{equation}

An essential ingredient in the thin-shell formalism is the extrinsic curvature, or second fundamental form, which is defined as $K_{ij}=n_{{(}\mu;\nu{)}}e^{\mu}_{(i)}e^{\nu}_{(j)}$,
where $n_{\mu}$ is the unit normal 4-vector (\ref{normalSigma}) to the surface $\Sigma$, and $e^{\mu}_{(i)}$ are the components of the holonomic basis of vectors tangent to $\Sigma$. Thus,  in terms of the above quantities, the extrinsic curvature can be expressed in the more tractable form:
\begin{equation}
    K_{ij}^{\pm} = -n_{\mu}\left(\frac{\partial^{2}x^{\mu}}{\partial\xi^{i}\partial\xi^{j}} + \Gamma^{\mu\pm}_{\ \alpha\beta}\frac{\partial x^{\alpha}}{\partial\xi^{i}}\frac{\partial x^{\beta}}{\partial\xi^{j}}\right).
\end{equation}

A quick calculation yields the $K_{\theta\theta}^{\pm}$ component, where the mixed tensor is given~by:
\begin{equation}
K^{\theta\pm}_{\ \theta} = g^{\theta\theta}\,K_{\theta\theta}^{\pm} = \pm\frac{1}{R}\sqrt{\gpm +\dot{R}^{2}}.
\end{equation}

A lengthy calculation yields the $K^{\tau\pm}_{\ \tau}$ component, but we make use of the formalism discussed in~\cite{Visser:1995cc}, which is rather pedagogical. 
To this effect, note that we have
\begin{eqnarray}
K^{\pm}_{\tau\tau} &=& K^{\pm}_{\mu\nu}U^{\mu}U^{\nu} 
= \nabla^{\pm}{}_{(\mu} \; n_{\nu)} U^{\mu}U^{\nu} 
	\nonumber  \\
&=&  \left[\frac{1}{2}\left(\nabla^{\pm}_{\mu}n_{\nu}+\nabla^{\pm}_{\nu}n_{\mu}\right)\right]U^{\mu}U^{\nu} = \nabla^{\pm}_{\mu}n_{\nu}U^{\mu}U^{\nu}.
\end{eqnarray}
Taking into account $K^{\tau\pm}_{\ \tau} = -K^{\pm}_{\tau\tau}$, we therefore have the following:
\begin{equation}
K^{\tau\pm}_{\ \tau} = -\left(\nabla^{\pm}_{\mu}n_{\nu}\right)U^{\mu}U^{\nu} 
= +U^{\mu}n_{\nu}\left(\nabla^{\pm}_{\mu}U^{\nu}\right)
= n_{\nu}\left(U^{\mu}\nabla^{\pm}_{\mu}U^{\nu}\right) 
= n_{\nu}A^{\nu}_{\pm}, 
\end{equation}
where $A^{\nu}_{\pm}$ is the $4$--acceleration of the throat. 
Spherical symmetry implies that $A^{\nu}_{\pm}\propto n^{\nu}$, i.e. $A^{\nu}_{\pm} = \abs{A_{\pm}} n^{\nu}$. 
Therefore:
\begin{equation}
K^{\tau\pm}_{\ \tau} = \left( n_{\nu} \abs{A_{\pm}} \right) n^{\nu} = \abs{A_{\pm}}.
\end{equation}
That is, $K^{\tau\pm}_{\ \tau}$ is simply equal to the magnitude of the $4$--acceleration of the throat.

The underlying bulk geometry possesses a Killing vector $k^{\mu} = \left(\partial_{t}\right)^{\mu} = \left(1, 0, 0, 0\right)^\mu$. 
Lowering the index on this Killing vector, we obtain (calculating \emph{at the throat} where $r=R(\tau)$)
\begin{equation} 
k_{\mu} = \left(-\left[\g\right], 0, 0, 0\right).
\end{equation}
We now examine the quantity $\frac{d}{d\tau}\left(k_{\mu}U^{\mu}\right)$, which we can compute in two different ways to obtain the magnitude of the $4$--acceleration as a function of $R$, its first and second derivatives, $a$ and $m$:

\begin{itemize}
    \item First calculation (employing Killing's equation):   
    \begin{eqnarray}
        \dv{\tau}\left(k_{\mu}U^{\mu}\right) 
        &=& U^{\nu}\nabla_{\nu}\left(k_{\mu}U^{\mu}\right)
           = \left(\nabla^{\pm}_{\nu}k_{\mu}\right)U^{\mu}U^{\nu}+k_{\mu}\dv{U^{\mu}}{\tau} 
        \nonumber \\
        && \nonumber \\
        &=& k_{\mu}\dv{U^{\mu}}{\tau} = k_{\mu}A^{\mu}_{\pm} 
        = k_{\mu}\vert A_{\pm}\vert n^{\mu} = \vert A_{\pm}\vert\left(k_{\mu}n^{\mu}\right) 
        \nonumber \\
        && \nonumber \\
        &=& \mp \vert A_{\pm}\;  \vert \dot{R}. \label{ktt1}
    \end{eqnarray}
\item Second calculation:
    \begin{eqnarray}
        \dv{\tau}\left(k_{\mu}U^{\mu}\right) &=& \dv{\tau}\left(k_{t}U^{t}\right) 
        = -\dv{\tau}\left[\sqrt{\g +\dot{R}^{2}}\right] \nonumber \\
        &=& -\frac{\dot{R}\left[\ddot{R}+\frac{m\,\e^{-\frac{a}{R}}}{R^{2}}\left(1-\frac{a}{R}\right)\right]}{\sqrt{\g +\dot{R}^{2}}}. \label{ktt2}
    \end{eqnarray}
\end{itemize}

Comparing equations \eqref{ktt1} and \eqref{ktt2}, we obtain
\begin{equation}
\mp\vert A_{\pm}\vert\dot{R} = -\frac{\dot{R}\left[\ddot{R}+\frac{m\,\e^{-\frac{a}{R}}}{R^{2}}\left(1-\frac{a}{R}\right)\right]}{\sqrt{\g +\dot{R}^{2}}},
\end{equation}
and so
\begin{equation}
\quad K^{\tau\pm}_{\ \tau} = \vert A_{\pm}\vert = \pm \left[\frac{\ddot{R} + \frac{m_{\pm}\e^{-\frac{a_{\pm}}{R}}}{R^{2}}\left(1-\frac{a_{\pm}}{R}\right)}{\sqrt{\gpm +\dot{R}^{2}}}\right].
\end{equation}

In summary, the extrinsic curvature components are given by
\begin{eqnarray}
K\indices{^\theta_\theta^\pm} &=& K\indices{^\phi_\phi^\pm} = \pm \frac{1}{R} \sqrt{\gpm + \dot{R}^2},		\label{curvtheta} 
	\\[10pt]
K\indices{^\tau_\tau^\pm} &=& \pm \left[ \frac{m_\pm \, \e^{-a_\pm/R} (R-a_\pm) + R^3 \ddot{R}}{R^3 \sqrt{\gpm+\dot{R}^2}} \right] \,,
\label{curvtau}
\end{eqnarray}
respectively.

\subsection{Surface stress--energy}

For our thin--shell analysis, the extrinsic curvature need not be continuous across the junction boundary \( \Sigma \).
Thus, we denote the discontinuity by \( \kappa_{ij} = K_{ij}^+ - K_{ij}^- \).
The surface stress--energy tensor on \( \Sigma \), \( S\indices{^i_j} \), can be calculated via the Lanczos equations:
\begin{equation}
S\indices{^i_j} = -\frac{1}{8\pi} \left( \kappa\indices{^i_j} - \delta\indices{^i_j} \kappa\indices{^k_k} \right).
\end{equation}
Due to spherical symmetry, the discontinuity can be represented by a diagonal matrix: \( \kappa\indices{^i_j} = \text{diag}(\kappa\indices{^\tau_\tau}, \kappa\indices{^\theta_\theta}, \kappa\indices{^\phi_\phi}) \), and so the surface stress--energy tensor simply reduces to \( S\indices{^i_j} = \text{diag}(-\sigma, \P, \P) \), where \( \sigma \) is the surface energy density and \( \P \) is the surface pressure.
Thus, with \( \kappa\indices{^k_k} = \kappa\indices{^\tau_\tau} + 2\kappa\indices{^\theta_\theta} \), the Lanczos equations imply:

\begin{align}
\sigma &= -\frac{1}{4\pi}\kappa\indices{^\theta_\theta},		\\
\P &= \frac{1}{8\pi}(\kappa\indices{^\tau_\tau} + \kappa\indices{^\theta_\theta}).
\end{align}

Using the extrinsic curvature components given in Eq's.~\eqref{curvtheta} and~\eqref{curvtau}, the surface stress--energy at the junction throat \( \Sigma \) is finally found to be:
\begin{eqnarray}
\sigma = -\frac{1}{4\pi R} \left[ \sqrt{\gp + \dot{R}^2} \, + \ \sqrt{\gm + \dot{R}^2} \right],	\label{eq;surfE}
\end{eqnarray}
\begin{eqnarray}
\P &=& \frac{1}{8\pi R} \left[ \frac{1+\dot{R}^2 + R\ddot{R} - \frac{m_+\,\e^{-a_+/R}}{R^2}(R+a_+)}{\sqrt{\gp+\dot{R}^2}} \right.	\notag \\
&&\hspace{4cm}+ \left. \frac{1+\dot{R}^2 + R\ddot{R} - \frac{m_-\,\e^{-a_-/R}}{R^2}(R+a_-)}{\sqrt{\gm+\dot{R}^2}} \right]. 	
\label{eq;surfP}
\end{eqnarray}
It can be seen from equation \eqref{eq;surfE} that negative energy is needed to keep the wormhole throat open, implying that exotic matter would be required.

An important ingredient explored in recent work~\cite{Garcia:2011aa,Lobo:2020kxn} is the potential presence of an additional energy flux term, which arises from the conservation identity. This identity is obtained by combining the second contracted Gauss--Codazzi equation (or the ``ADM" constraint) $G_{\mu \nu}\,e^{\mu}_{(i)}n^{\nu}=K^j_{i|j}-K,_{i}$ with the Lanczos equations, and is given by $S^{i}_{j|i}=-\left[T_{\mu \nu}e^{\mu}_{(j)}n^{\nu}\right]^+_-$.
The momentum flux term in the right hand side corresponds to the net discontinuity in the momentum which impinges on the shell. Note that for the present geometry, this flux term vanishes:
\begin{eqnarray}
    \left[T_{\mu\nu}e^{\mu}_{(\tau)}n^{\nu}\right]^{+}_{-} 
    &=& \left[T_{\mu\nu}U^{\mu}n^{\nu}\right]^{+}_{-} 
		\nonumber \\    
    &=&  \left[\pm\left(-T_{t}{}^{t}+T_{r}{}^{r}\right)
    \frac{\dot{R}\sqrt{1-\frac{2m_{\pm}\e^{-a_{\pm}/R}}{R}+\dot{R}^{2}}}
    {1-\frac{2m_{\pm}\e^{-a_{\pm}/R}}{R}}
    \right]^{+}_{-} = 0 \,,
\end{eqnarray}
where $T_{t}{}^{t}=-\rho$ and $T_{r}{}^{r}=p_r$, and Eq. (\ref{rhopr}) yields $-T_{t}{}^{t}+T_{r}{}^{r}=\rho+p_r=0$.

Thus, the conservation identity finally provides $S^{i}_{\tau|i}=0=-\left[\dot{\sigma}+2\dot{a}(\sigma +{\cal P} )/a \right]$. That is:
\begin{equation}
\sigma'=-\frac{2}{a}\,(\sigma+{\cal P})
  \,.
\label{consequation2}
\end{equation}

\subsection{Stability analysis}\label{sec:stability}

\subsubsection{Equation of motion}

In order to force stability constraints on the mass of the thin-shell, 
$m_{s}(R)$, let us consider the thin--shell equation of motion; and write it in the form $\frac{1}{2}\dot{R}^{2}+V(R) = 0$. 
To obtain an explicit expression for the potential $V(R)$, taking into account $m_{s}(R) = 4\pi R^{2}\sigma(R)$, we rearrange  Eq.~(\ref{eq;surfE}) to derive:
\begin{equation}
    V(R) = -\frac{1}{2}\dot{R}^{2} = \frac{1}{2}\left\lbrace 1+\frac{\bar{\Delta}(R)}{R}-\left[\frac{m_{s}(R)}{2R}\right]^{2}-\left[\frac{\Delta(R)}{m_{s}(R)}\right]^{2}\right\rbrace.
\end{equation}
Here $\bar{\Delta}(R)$ and $\Delta(R)$ are defined as:
\begin{eqnarray}
    \bar{\Delta}(R) &=& m_{+}\e^{-a_{+}/R}+m_{-}\e^{-a_{-}/R} \,, \qquad 
    \Delta(R) = m_{+}\e^{-a_{+}/R}-m_{-}\e^{-a_{-}/R}\,.
     \nonumber 
\end{eqnarray}

Having obtained this explicit form for $V(R)$, we may now recast the surface energy density $\sigma$ as a function of the effective potential:
\begin{equation}\label{sigmaV(R)}
    \sigma(R) = -\frac{1}{4\pi R} \left[ \sqrt{\gp - 2V(R)} \, + \ \sqrt{\gm - 2V(R)} \right] \ .
\end{equation}

\subsubsection{Linearized equation of motion}

Let us assume there exists some static solution at $R=R_{0}$, and linearize around it accordingly. The equation of motion is $\frac{1}{2}\dot{R}^{2} + V(R) = 0$, which also directly yields that $\ddot{R} = -V'(R)$, and if we analyse a second--order Taylor series expansion of $V(R)$ about $R_{0}$ we obtain the following:
\begin{equation}
    V(R) = V(R_{0})+V'(R_{0})\,(R-R_{0})+\frac{1}{2}V''(R_{0})\,(R-R_{0})^{2}+O[(R-R_{0})^{3}] \,.
\end{equation}
Various simplifications ensue due to our solution being \emph{static}, namely, 
$\dot{R}_{0} = \ddot{R}_{0} = 0$ and $V'(R_{0}) = -\ddot{R}_{0} = 0$.
\enlargethispage{50pt}

Thus, our Taylor series for $V(R)$ reduces to:
\begin{equation}
    V(R) = \frac{1}{2}V''(R_{0})(R-R_{0})^{2}+O[(R-R_{0})^{3}] \ .
\end{equation}
Now, the condition for our solution at $R_{0}$ to be stable is that $V(R_{0})$ is a local minima; \emph{i.e.} $V''(R_{0}) > 0$. Given our form for $\sigma$ as a function of $V(R)$ in Eq.~(\ref{sigmaV(R)}), we may now use this condition, along with $V(R_{0})=V'(R_{0})=0$, to force stability constraints on the mass of the thin--shell. It is in fact preferable to consider the effect of these constraints on the \emph{dimensionless} quantity $\left[{m_{s}(R)}/{R}\right]$, rather than on $m_{s}(R)$ itself. 

In all generality we have the following:
\begin{eqnarray}
    \frac{m_{s}(R)}{R} = 4\pi\sigma(R)R &=& -\Bigg[ \sqrt{\gp - 2V(R)} \, 
		\nonumber \\    
    && 
  \hspace{1.5cm} +  \sqrt{\gm - 2V(R)} \Bigg] \,,
\end{eqnarray}
\begin{eqnarray}    
    \left[\frac{m_{s}(R)}{R}\right]' = -\left[\frac{\frac{m_{+}\e^{-a_{+}/R}(R-a_{+})}{R^{3}}-V'(R)}{\sqrt{\gp - 2V(R)}} + \frac{\frac{m_{-}\e^{-a_{-}/R}(R-a_{-})}{R^{3}}-V'(R)}{\sqrt{\gm - 2V(R)}}\right] , \;\;
    \end{eqnarray}
\begin{eqnarray}
    \left[\frac{m_{s}(R)}{R}\right]'' &=& \frac{\left[\frac{m_{+}\e^{-a_{+}/R}}{R^{2}}\left(1-\frac{a_{+}}{R}\right)-V'(R)\right]^{2}}{\left[1-\frac{2m_{+}\e^{-a_{+}/R}}{R}-2V(R)\right]^{\frac{3}{2}}} - \frac{\frac{m_{+}a_{+}\e^{-a_{+}/R}}{R^{4}}\left(4-\frac{a_{+}}{R}\right)-V''(R)}{\sqrt{1-\frac{2m_{+}\e^{-a_{+}/R}}{R}-2V(R)}} \nonumber \\
    && \nonumber \\
    && + \frac{\left[\frac{m_{-}\e^{-a_{-}/R}}{R^{2}}\left(1-\frac{a_{-}}{R}\right)-V'(R)\right]^{2}}{\left[1-\frac{2m_{-}\e^{-a_{-}/R}}{R}-2V(R)\right]^{\frac{3}{2}}} - \frac{\frac{m_{-}a_{-}\e^{-a_{-}/R}}{R^{4}}\left(4-\frac{a_{-}}{R}\right)-V''(R)}{\sqrt{1-\frac{2m_{-}\e^{-a_{-}/R}}{R}-2V(R)}} \,.
    \nonumber \\
\end{eqnarray}

\subsubsection{Master equations}
Applying the stability constraints to these equations, we see that in order to have a stable solution at $R_{0}$ the thin-shell mass $m_s(R)$  must satisfy the following:

\begin{eqnarray}\label{stabilityineq0}
    \frac{m_{s}(R_{0})}{R_{0}} &=& -\left[\sqrt{1-\frac{2m_{+}\e^{-a_{+}/R_{0}}}{R_{0}}} + \sqrt{1-\frac{2m_{-}\e^{-a_{-}/R_{0}}}{R_{0}}}\right] \,,
    \end{eqnarray}
    \begin{eqnarray}
    \left[\frac{m_{s}(R_{0})}{R_{0}}\right]' &=& -\left[\frac{m_{+}\e^{-a_{+}/R_{0}}(R_{0}-a_{+})}{R_{0}^{3}\sqrt{1-\frac{2m_{+}\e^{-a_{+}/R_{0}}}{R_{0}}}} + \frac{m_{-}\e^{-a_{-}/R_{0}}(R_{0}-a_{-})}{R_{0}^{3}\sqrt{1-\frac{2m_{-}\e^{-a_{-}/R_{0}}}{R_{0}}}}\right] \,,
    \end{eqnarray}
     \begin{eqnarray}
    \left[\frac{m_{s}(R_{0})}{R_{0}}\right]'' &\geq& \frac{\left[\frac{m_{+}\e^{-a_{+}/R_{0}}}{R_{0}^{2}}\left(1-\frac{a_{+}}{R_{0}}\right)\right]^{2}}{\left[1-\frac{2m_{+}\e^{-a_{+}/R_{0}}}{R_{0}}\right]^{\frac{3}{2}}} - \frac{\frac{m_{+}a_{+}\e^{-a_{+}/R_{0}}}{R_{0}^{4}}\left(4-\frac{a_{+}}{R_{0}}\right)}{\sqrt{1-\frac{2m_{+}\e^{-a_{+}/R_{0}}}{R_{0}}}} \nonumber \\
    && \nonumber \\
    && + \frac{\left[\frac{m_{-}\e^{-a_{-}/R_{0}}}{R_{0}^{2}}\left(1-\frac{a_{-}}{R_{0}}\right)\right]^{2}}{\left[1-\frac{2m_{-}\e^{-a_{-}/R_{0}}}{R_{0}}\right]^{\frac{3}{2}}} - \frac{\frac{m_{-}a_{-}\e^{-a_{-}/R_{0}}}{R_{0}^{4}}\left(4-\frac{a_{-}}{R_{0}}\right)}{\sqrt{1-\frac{2m_{-}\e^{-a_{-}/R_{0}}}{R_{0}}}} \ .
    \label{stabilityineq}
\end{eqnarray}
This final inequality gives us the stability regions for the thin--shell wormhole for various cases of the parameters $m_{\pm}$ and $a_{\pm}$.

\section{Examples}

Let us now analyse some of the more interesting specific sub--cases by fixing the parameters $a_{\pm}$ and $m_{\pm}$ and examining the corresponding stability criteria implied by Eq.~(\ref{stabilityineq}).

\subsection{Symmetrically vanishing $a$ parameter; asymmetric mass $m_+\neq m_-$.}

In the bulk spacetime we know that $a=0$ corresponds to the usual Schwarzschild solution. To fix $a_{+}=a_{-}=0$ in the wormhole construction while allowing asymmetric masses $m_{-}\neq m_{+}$ is to perform the thin--shell surgery exterior to two Schwarzschild spacetimes with distinct masses. By now, this particular thin--shell construction is rather well--known; see~\cite{Garcia:2011aa, Lobo:2020kxn}. For the purposes of plotting the stability regions we define a dimensionless form for the stability constraint as follows.
First note that the Eq.~(\ref{stabilityineq}) reduces to:
\begin{equation}
    R_{0}^{2}\left[\frac{m_{s}(R_{0})}{R_{0}}\right]''\geq F_{1}(R_{0}, m_{\pm}) = \frac{m_{+}^{2}}{R_{0}^{2}\left(1-\frac{2m_{+}}{R_{0}}\right)^{\frac{3}{2}}} + \frac{m_{-}^{2}}{R_{0}^{2}\left(1-\frac{2m_{-}}{R_{0}}\right)^{\frac{3}{2}}} \ .
\end{equation}
Then, for the purposes of plotting the full domain of $R_{0}$, we shall consider the dimensionless definitions $x=\frac{2m_{+}}{R_{0}} \ , y=\frac{2m_{-}}{R_{0}}$, so that the parameters $x$ and $y$ lie in the ranges $0<x<1$ and  $0<y<1$, respectively.
Hence:
\begin{equation}\label{F1}
    F_{1}(x, y) = \frac{1}{4}
    \left[\frac{x^{2}}{\left(1-x\right)^{\frac{3}{2}}}+\frac{y^{2}}{\left(1-y\right)^{\frac{3}{2}}}\right] \ .
\end{equation}

\begin{figure}[!htb]
\begin{center}
\includegraphics[scale=0.65]{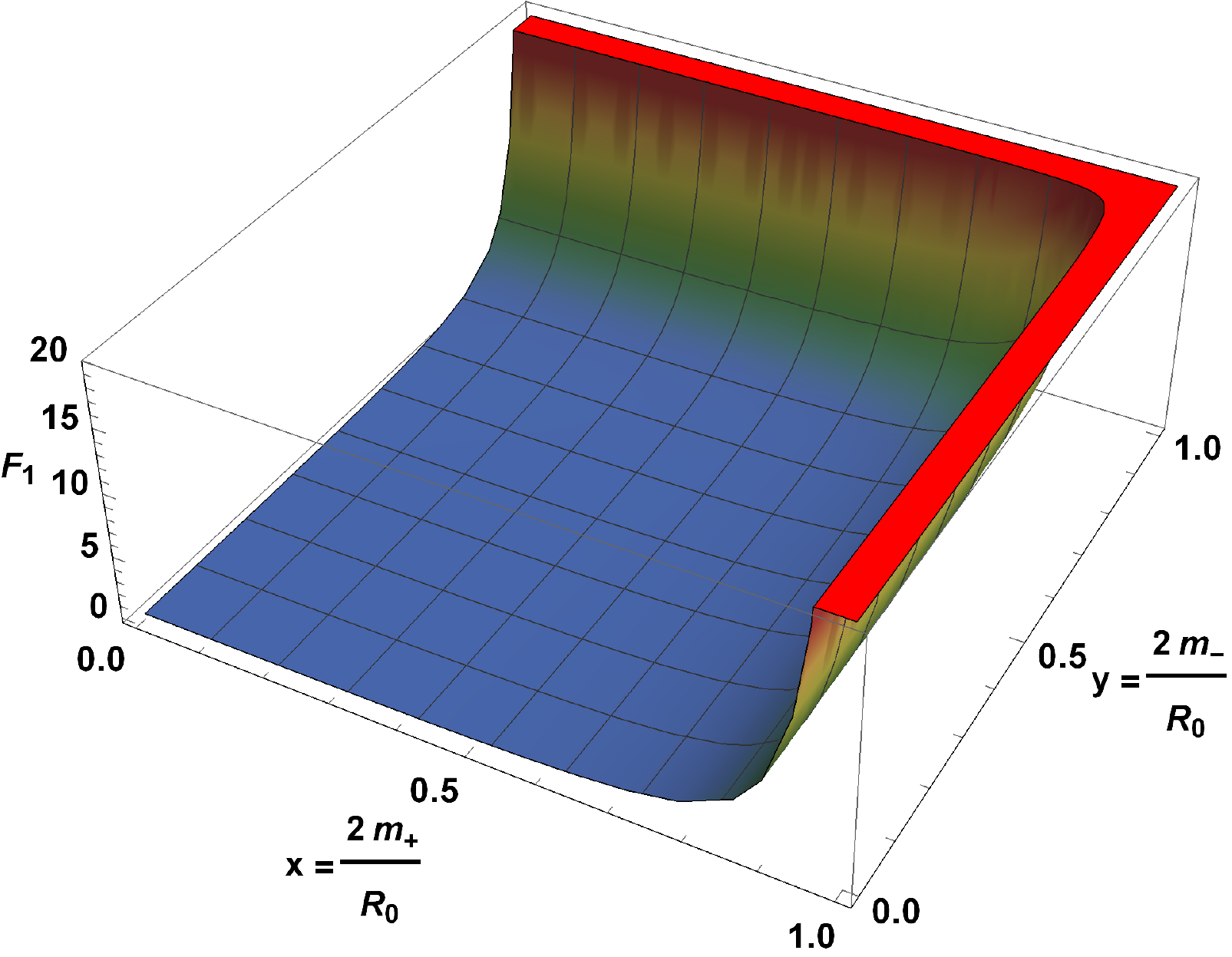}
\end{center}
\caption{Stability analysis for the $a_{\pm}=0$ case, which reduces to Schwarzschild surgery. The stability region lies above the surface $F_{1}(x, y)$, as given explicitly by Eq.~(\ref{F1}). The red region indicates where this function departs the specified range for $z$, and we can see that $F_{1}$ is increasing \emph{very steeply} within this region, as $x\rightarrow 1$ and/or $y\rightarrow 1$. Note that we chop off the plot  vertically once $F_1(x,y)=20$.
}
\label{F:1}
\end{figure}

We see from Figure~\ref{F:1} that large stability regions exist for low values of $x$ and $y$, corresponding to $R_{0}\gg 2m_{\pm}$, while as $R_{0}\rightarrow 2m_{\pm}$ the size of the stability regions decreases steeply as we near the respective horizons.

The special case of equal masses $m_+=m_-$ simply corresponds to the diagonal $x=y$ in Figure~\ref{F:1}.
Before proceeding to the next case of interest it is worth noting that, since our construction is formed from a spacetime which is strictly Minkowski in the $m\rightarrow 0$ limit, the case of symmetrically vanishing $m_{\pm}=0$ trivially reduces to Minkowski surgery. 
This corresponds to $x=0=y$ and $F_1(0,0)=0$. Thence in this specific situation the stability criterion simply reduces to
\begin{equation}
    \left[\frac{m_{s}(R_{0})}{R_{0}}\right]''\geq 0  .
\end{equation}
Similar logic is applied to the case for asymmetric vanishing of parameters, say (without loss of generality) for $m_+>0$ while $m_{-}=0$, as we are simply stitching Schwarzschild with Minkowski. This corresponds to $y=0$ but with $x>0$, and is represented by the $x$-axis in Figure~\ref{F:1}.

\subsection{Mirror symmetry: Both $m_+=m_-$ and $a_+=a_-$}

For the specific case of mirror symmetry, let us fix both  $m_{+}=m_{-}=m$ as well as $a_{+}=a_{-}=a$. For this case the stability condition reduces to:
\begin{equation}
    R_{0}^{2}\left[\frac{m_{s}(R_{0})}{R_{0}}\right]'' 
    \geq F_{2}(R_{0}, m, a) 
    = 2\left\lbrace\frac{\left[m\,\e^{-a/R_{0}}\left(1-\frac{a}{R_{0}}\right)\right]^{2}}
    {R_{0}^{2}\left[1-\frac{2m\,\e^{-a/R_{0}}}{R_{0}}\right]^{\frac{3}{2}}} 
    - \frac{ma\,\e^{-a/R_{0}}\left(4-\frac{a}{R_{0}}\right)}
    {R_{0}^{2}\sqrt{1-\frac{2m\,\e^{-a/R_{0}}}{R_{0}}}}\right\rbrace \ .
\end{equation}
In this case, we consider the two dimensionless parameters $x=\frac{2m}{R_{0}}\; \e^{-a/R_0}$ and $y=\frac{a}{R_{0}}$. 
Then the dimensionless function $F_{2}(x, y)$ is given by:
\begin{equation}\label{F2}
    F_{2}(x, y) = { x^2(1-y)^2\over2(1-x)^{3/2}} - {x y (4-y)\over(1-x)^{1/2}}.
\end{equation}
Notice that $x\in[0,1)$ to keep $F_{2}(x, y)$ real and finite. Furthermore, if the bulk spacetime contains horizons then $y\in(0,1]$; if the bulk spacetime is horizonless we are allowed to enter the region $y\in(1,\infty)$. Observe that the parameter $x$ has a natural directly physical interpretation in terms of the gravitational redshift $z$ of the throat as seen from spatial infinity:
\begin{equation}
1+z = {1\over\sqrt{1-x}} = {1\over \sqrt{ 1 - \frac{2m}{R_{0}}\; \e^{-a/R_0}}}.
\end{equation}
The point $(x,y)=(1,1)$ corresponds to the wormhole throat being located exactly at the degenerate horizon of an extremal bulk spacetime. 
The region $(x,y)\approx(1,1)$ corresponds to the wormhole throat being located near the almost degenerate horizon of a near-extremal bulk spacetime. 
It is easy to check that
\begin{equation}
\lim_{x\to1} F_{2}(x, y\neq 1) = +\infty; \qquad\qquad
\lim_{x\to1} F_{2}(x, y=1) = -\infty. 
\end{equation}

\begin{figure}[!htbp]
\begin{center}
\includegraphics[scale=0.55]{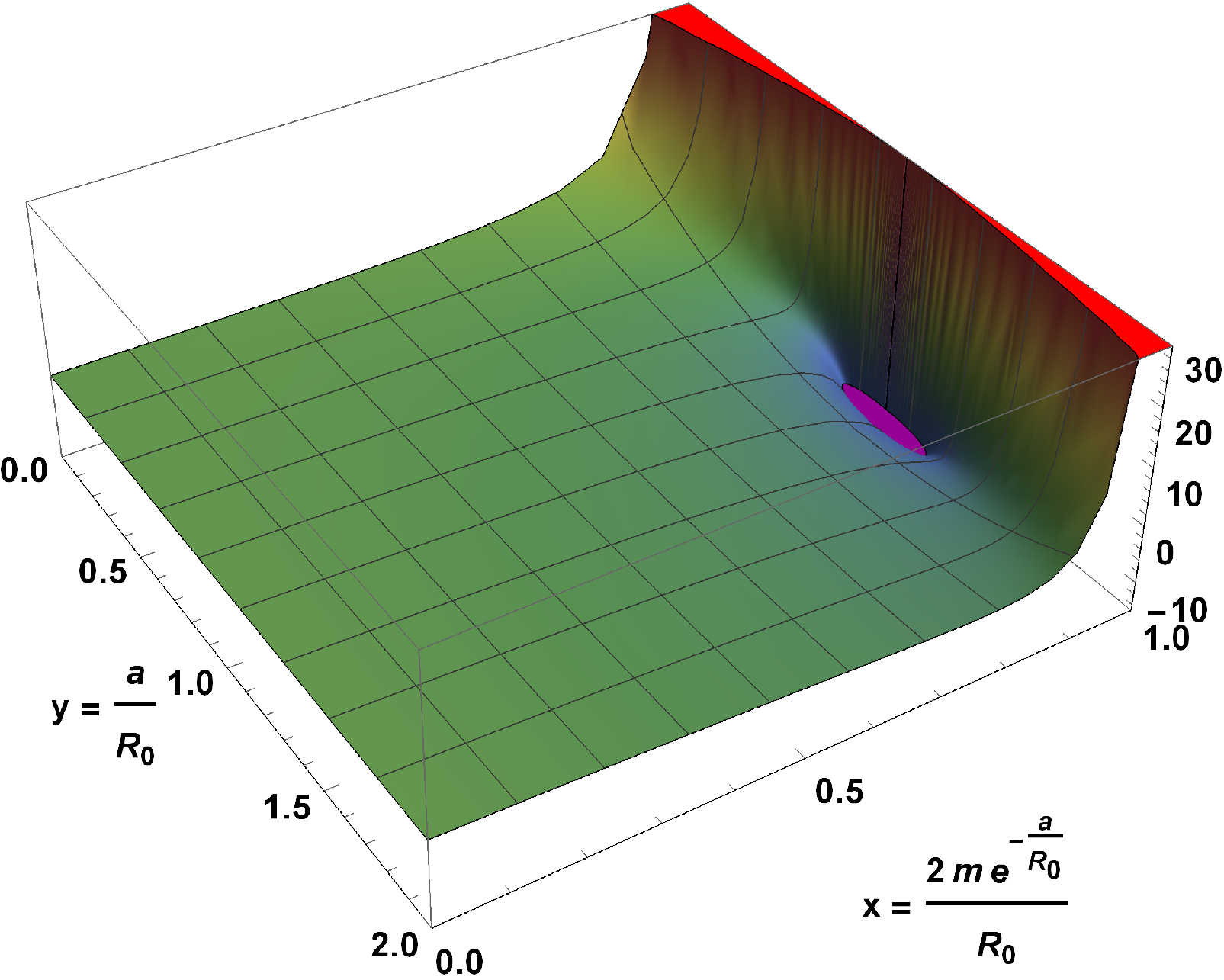}
\end{center}
\caption{Stability analysis for the case of perfect mirror symmetry; $a_{+}=a_{-}$, and $m_{+}=m_{-}$. The stability region lies above the surface $F_{2}(x, y)$, given explicitly by Eq.~(\ref{F2}). The red and purple regions indicates where this function departs the specified range.
Note that we chop the graph vertically at $F_2(x,y)=30$ and at $F_2(x,y)=-10$.
}
\label{F:2}
\end{figure}

\begin{figure}[!htbp]
\begin{center}
\includegraphics[scale=0.3]{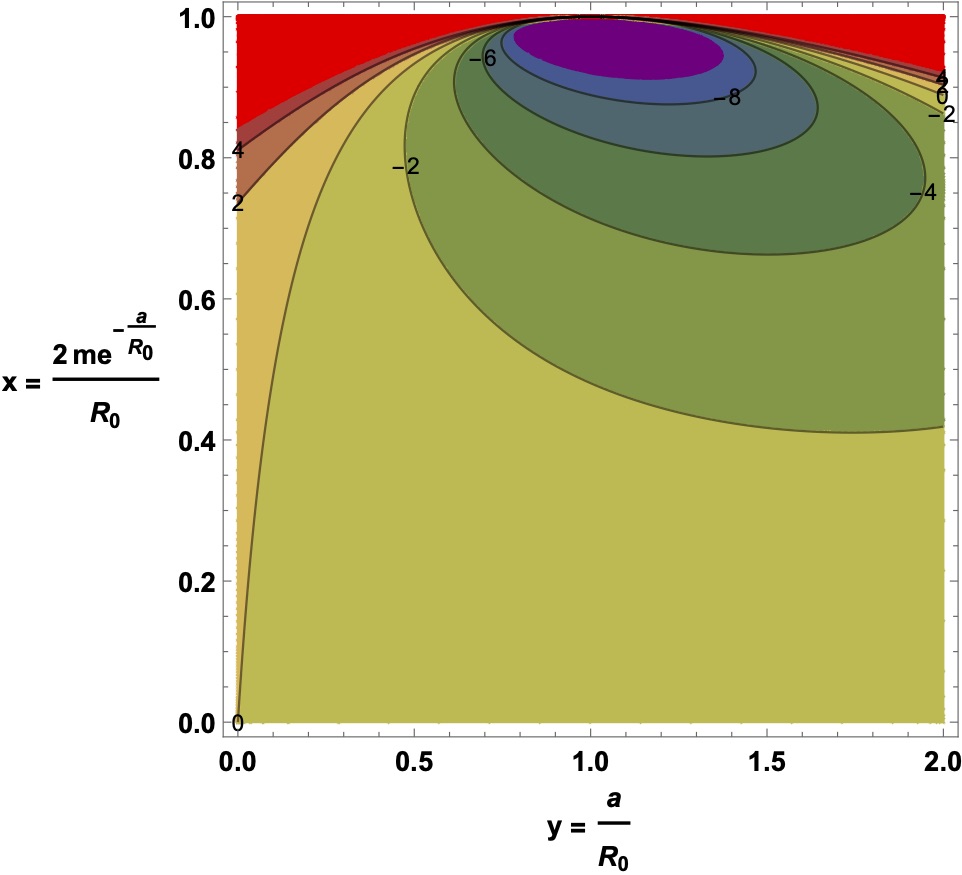}
\end{center}
\caption{Contour plot: Stability analysis for the case of perfect mirror symmetry.
The purple region indicates the `pit' where  $F_2(x,y)<-10$. 
The red region indicates the region of lesser sability where  $F_2(x,y)>30$.
}
\label{F:2b}
\end{figure}

Inspecting  Figures~\ref{F:2} and~\ref{F:2b} we observe relatively large stability regions.
An interesting feature of this plot is the presence of a  `pit' in the behaviour of $F_{2}(x,y)$ where the function is \emph{significantly} negative in the immediate vicinity of the extremal point $(x,y)=(1,1)$. 
This `pit' is a region which maximises the size of the stability region, and hence implies a preferred location for $R_{0}$ as a function of $m$ and $a$. 
\enlargethispage{40pt}

\begin{figure}[!htb]
\begin{center}
\includegraphics[scale=0.50]{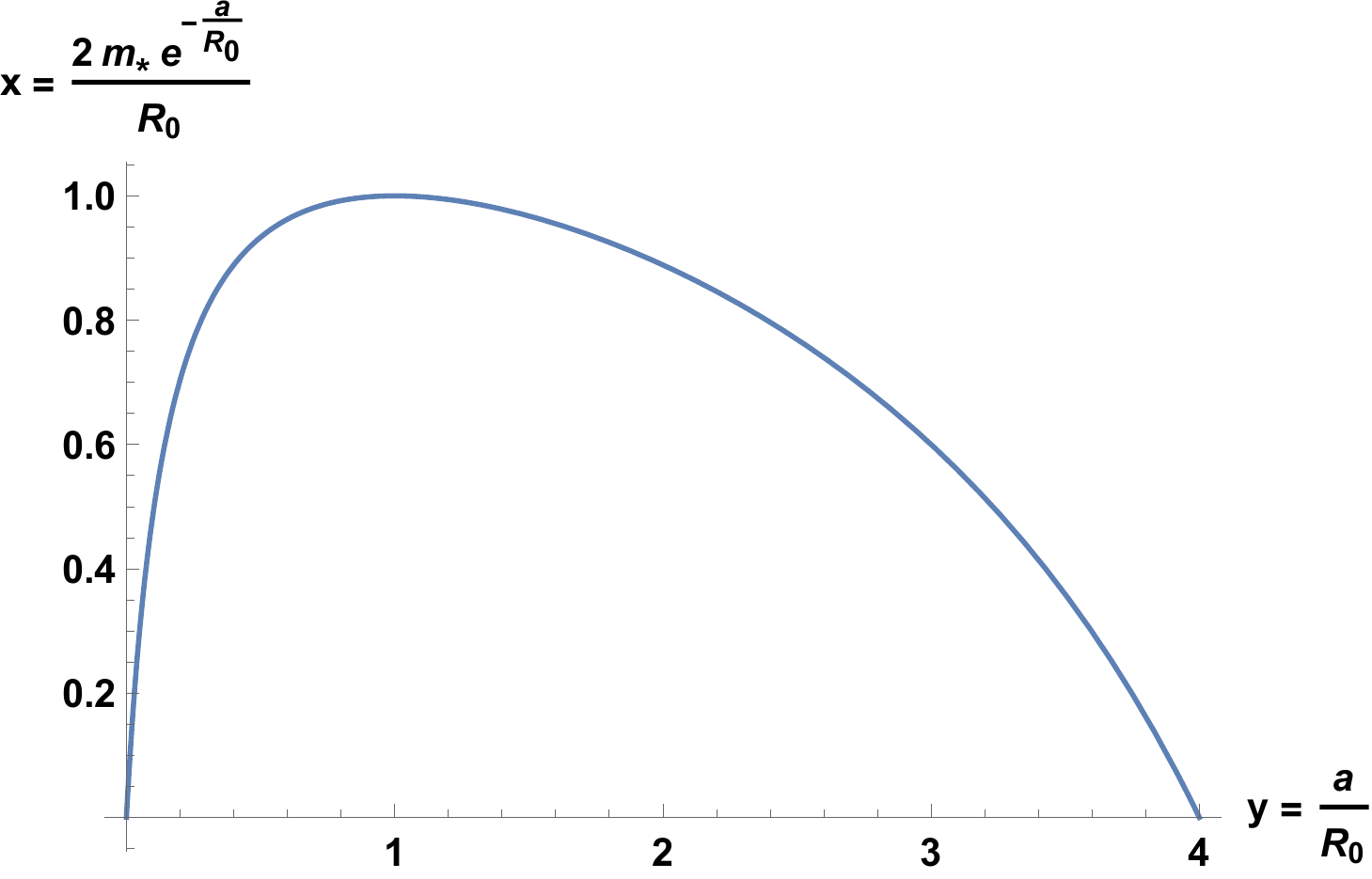}
\end{center}
\caption{Region in the $(x,y)$ plane where $F_2(x,y)$ flips sign.
}
\label{F:2c}
\end{figure}

\begin{figure}[!htb]
\begin{center}
\includegraphics[scale=0.50]{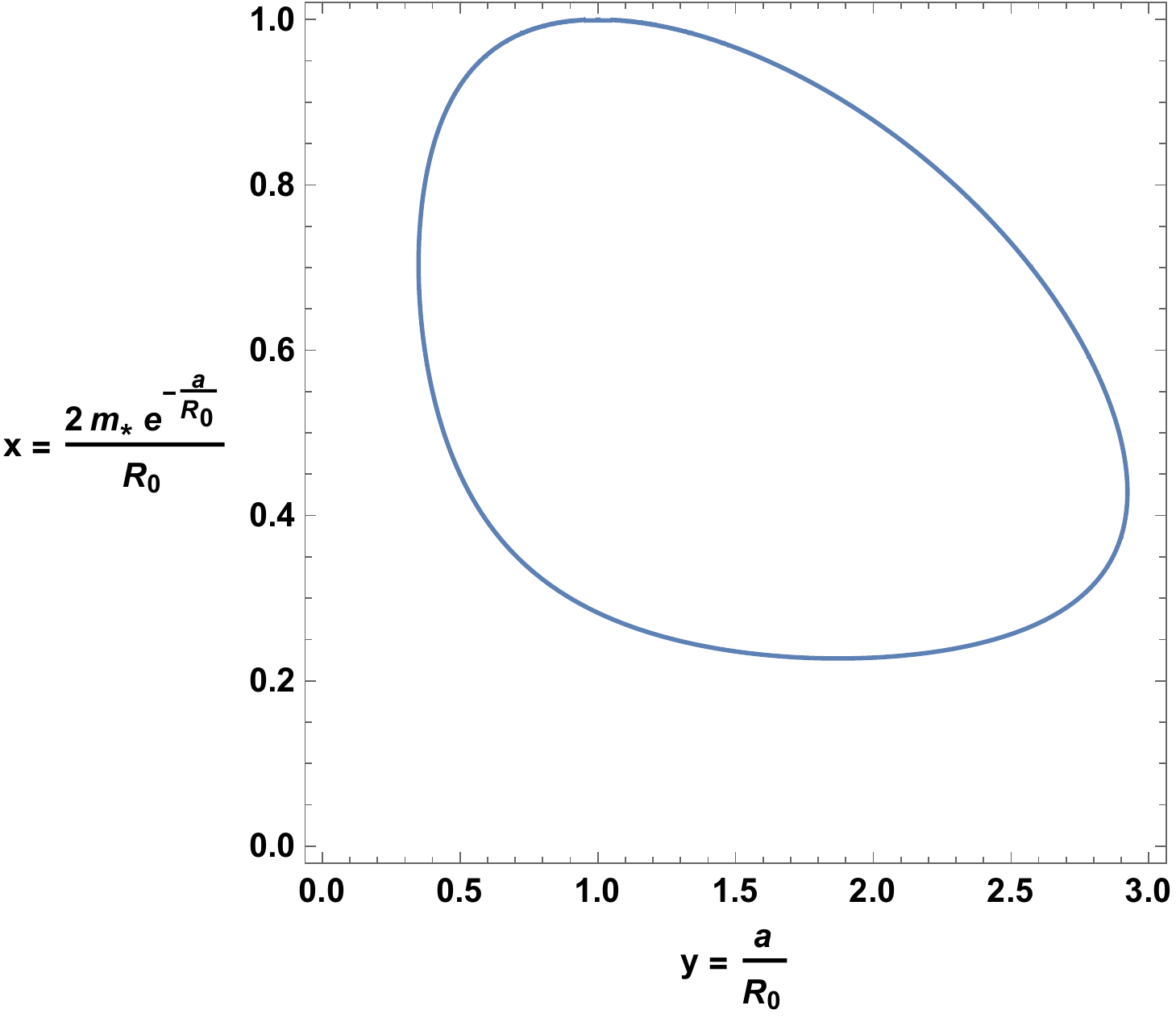}
\end{center}
\caption{Region in the $(x,y)$ plane where $F_2(x,y)<-1$.
}
\label{F:2d}
\end{figure}

The condition $F_{2}(x, y) = 0$, bounding the region where $F_{2}(x, y)$ changes sign, implicitly defines the curve 
\begin{equation}
x = {2y(4-y)\over1+6y-y^2}.
\end{equation}
In Figure~\ref{F:2c} we plot the boundary of this region where $F_{2}(x,y)$ changes sign.
Then in Figure~\ref{F:2d} we move deeper into the `pit' and plot the boundary of the region where $F_{2}(x,y)<-1$.

\clearpage

This `pit' in the stability plot is ultimately due to the wormhole throat getting close to where the extremal horizon would be in the bulk spacetime. It is actually a well-known phenomenon that having a wormhole throat get close to where a horizon would be in the bulk spacetime leads to interesting behaviour~\cite{small1,small2}.
In particular, we note that in this symmetric situation
\begin{eqnarray}\label{stabilityineq0b}
    \frac{m_{s}(R_{0})}{R_{0}} &=& -2\sqrt{1-\frac{2m\,\e^{-a/R_{0}}}{R_{0}}} 
    =  -2\sqrt{1-x}\,,
\end{eqnarray}
so that $x\approx 1$ corresponds to an arbitrarily small violation of the energy conditions~\cite{small1,small2}. In terms of the redshift of the throat
\begin{eqnarray}\label{stabilityineq0c}
    \frac{m_{s}(R_{0})}{R_{0}} &=& 
    -{2\over(1+z)^2}.
    \end{eqnarray}

\subsection{Specific asymmetry: $m_+\neq m_-$ while $a_+= a_-=a$}

Let us suppose $m_{+}\neq m_{-}$ while $a_{+}=a_{-}=a$. Hence we now have the case of surgery between two asymptotically Minkowski regular black holes with different masses but identical exponential suppression parameters. For a tractable analysis, 
let us define:
\begin{equation}
m_* = \max\{ m_+, m_-\}; \qquad 
\alpha = {\min\{m_+,m_-\} \over \max\{m_+,m_-\} } \leq 1,
\end{equation}

We may then re--express the stability condition of  Eq.~(\ref{stabilityineq}) as:\
\begin{eqnarray}
    R_{0}^{2}\left[\frac{m_{s}(R_{0})}{R_{0}}\right]'' &\geq & F_{3}(R_{0}, m_{*}, a, \alpha)    
		\nonumber \\ 
    &=& \frac{\left[\alpha m_{*}\e^{-a/R_{0}}\left(1-\frac{a}{R_{0}}\right)\right]^{2}}{R_{0}^{2}\left[1-\frac{2\alpha m_{*}\e^{-a/R_{0}}}{R_{0}}\right]^{\frac{3}{2}}} - \frac{\alpha m_{*}a\,\e^{-a/R_{0}}\left(4-\frac{a}{R_{0}}\right)}{R_{0}^{2}\sqrt{1-\frac{2\alpha m_{*}\e^{-a/R_{0}}}{R_{0}}}} \nonumber \\
    && \nonumber \\
    && + \frac{\left[m_{*}\e^{-a/R_{0}}\left(1-\frac{a}{R_{0}}\right)\right]^{2}}{R_{0}^{2}\left[1-\frac{2m_{*}\e^{-a/R_{0}}}{R_{0}}\right]^{\frac{3}{2}}} - \frac{m_{*}a\,\e^{-a/R_{0}}\left(4-\frac{a}{R_{0}}\right)}{R_{0}^{2}\sqrt{1-\frac{2m_{*}\e^{-a/R_{0}}}{R_{0}}}} \,,
\end{eqnarray}

Now define the two \emph{dimensionless} parameters, $x=\frac{2m_{*}}{R_{0}}\, \e^{-a/R_0}$ and  $y=\frac{a}{R_{0}}$,
so that the dimensionless function $F_{3}(x, y)$ takes the form
\begin{equation}
F_{3}(x, y) 
= \frac{\left[\alpha x (1-y)\right]^{2}}{4\left[1-\alpha x\right]^{\frac{3}{2}}} 
- \frac{\alpha x y (4-y)}{2\sqrt{1-\alpha x}} 
+ \frac{\left[x(1-y)\right]^{2}}{4\left[1-x\right]^{\frac{3}{2}}} 
- \frac{xy(4-y)}{2\sqrt{1-x}} \,.
\end{equation}
\enlargethispage{40pt}

\begin{figure}[htb!]
\begin{subfigure}{.5\textwidth}
\includegraphics[scale=0.450]{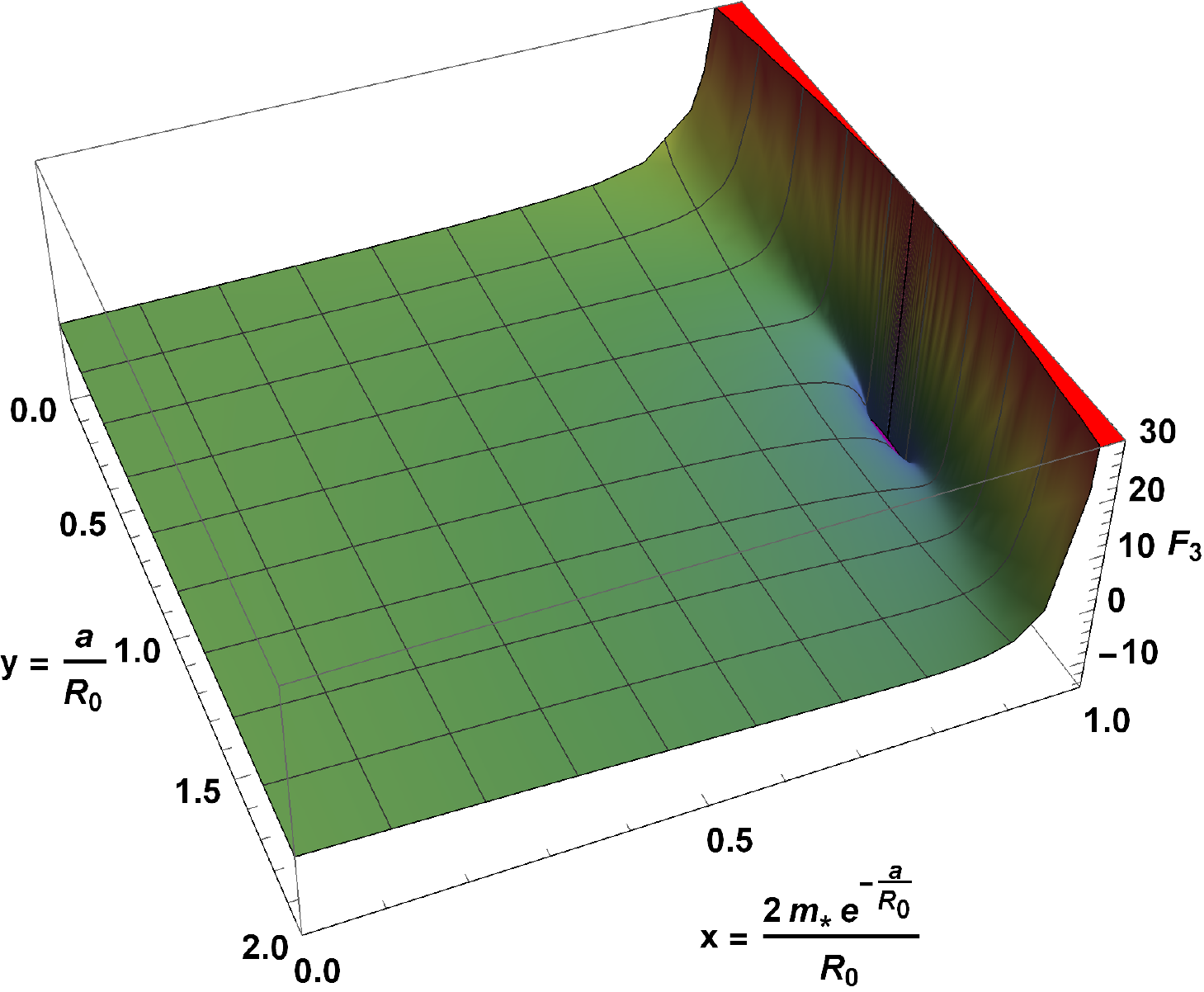}
\caption{$\alpha = 0.7$}
\end{subfigure}
\begin{subfigure}{.5\textwidth}
\includegraphics[scale=0.450]{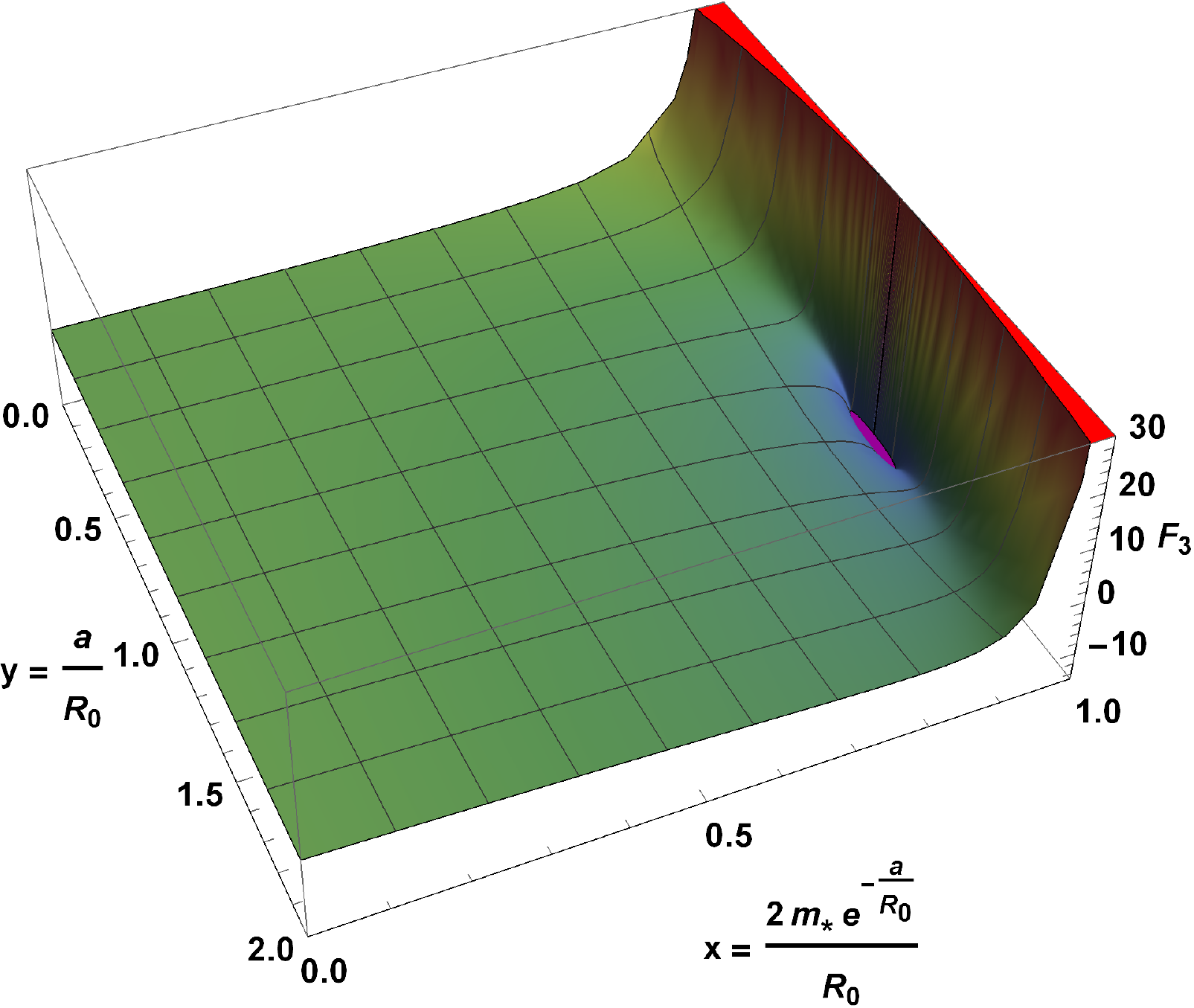}
\caption{$\alpha=0.9$}
\end{subfigure}
\caption{
Stability analysis for the specific asymmetry; $a_{+}=a_{-}=a$, while $m_{+}\neq m_{-}$. The stability region lies above the surface $F_{3}(x, y)$. The red and purple regions indicate where the function departs the specified range for $F_3(x,y)$.}
\label{F:3}
\end{figure}

\begin{figure}[htb!]
\begin{subfigure}{.5\textwidth}
\includegraphics[scale=0.215]{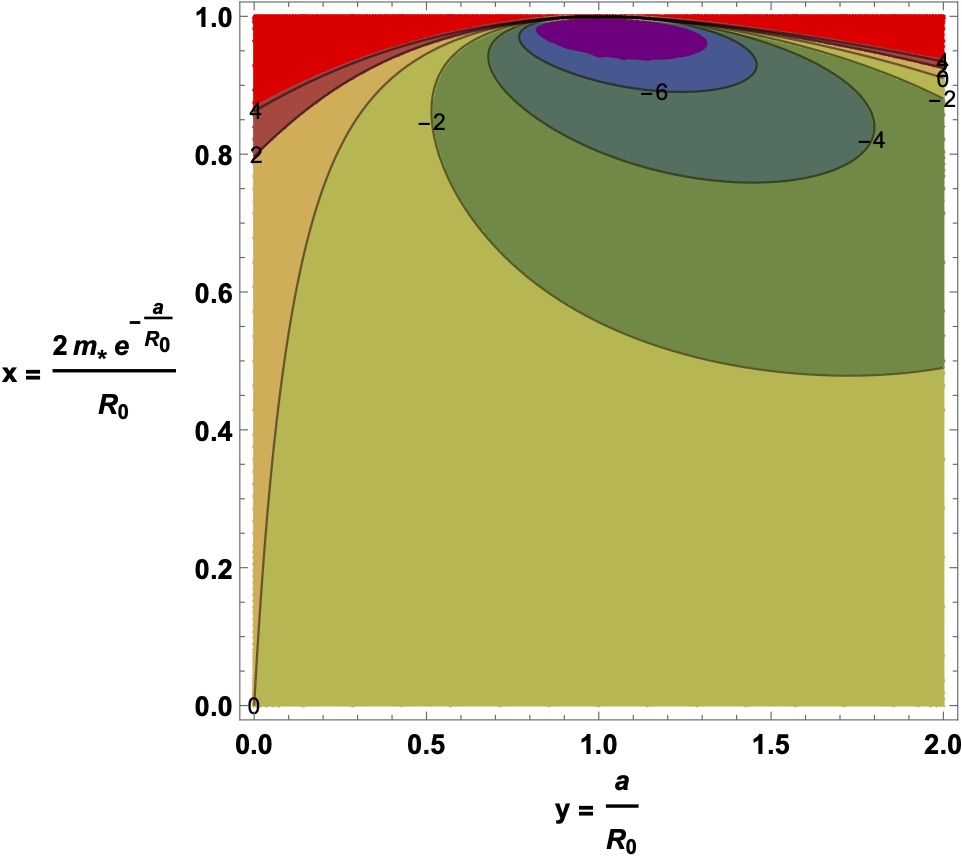}
\caption{$\alpha = 0.7$}
\end{subfigure}
\begin{subfigure}{.5\textwidth}
\includegraphics[scale=0.215]{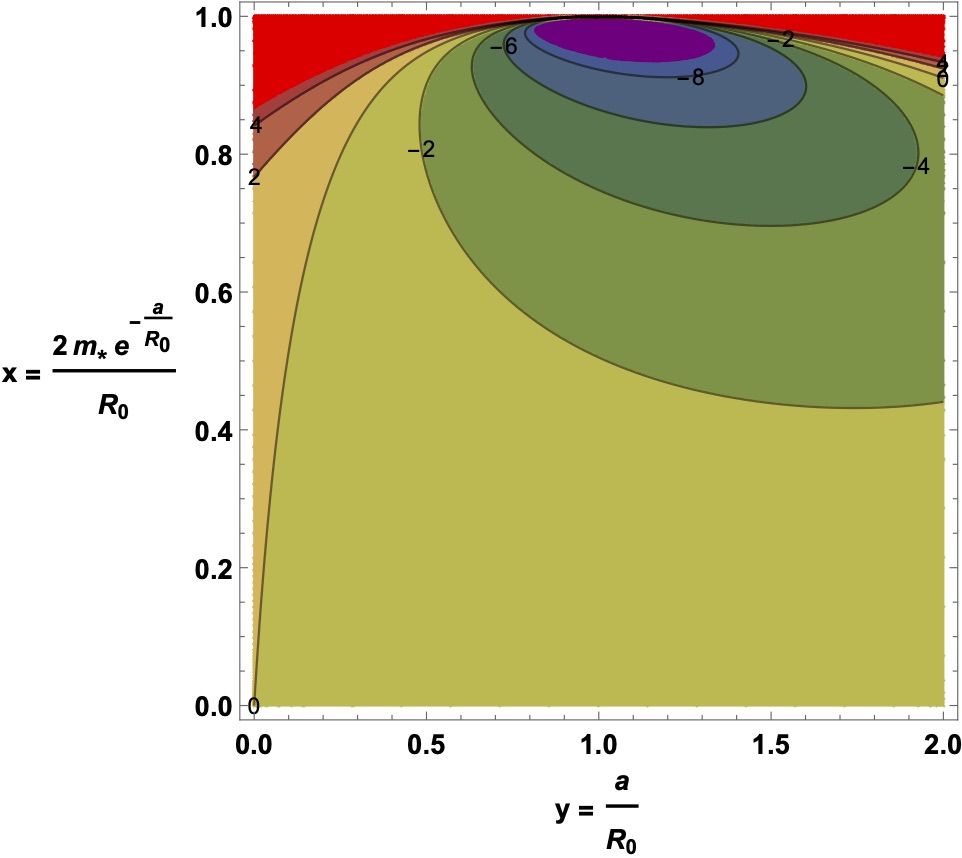}
\caption{$\alpha=0.9$}
\end{subfigure}
\caption{
Contour plots: Stability analysis for the specific asymmetry; $a_{+}=a_{-}=a$, while $m_{+}\neq m_{-}$. 
The purple region indicates `pit' where $F_3(x,y)<-10$.
The red region indicates region of lesser stability  where $F_3(x,y)$ is large and positive.}
\label{F:3b}
\end{figure}

Note that the argument of the square root on the denominator forces our $x$--parameter to be less than unity, otherwise $F_{3}(x,y)$ will become complex. We therefore have $0<x<1$, while $0<y\leq1$ if the bulk spacetimes have horizons, and $y\in(1,\infty)$ is allowed if the bulk spacetimes are horizonless. Since by construction $\alpha\leq 1$ it is easy to check that
\begin{equation}
\lim_{x\to1} F_{3}(x, y\neq 1) = +\infty; \qquad\qquad
\lim_{x\to1} F_{3}(x, y=1) = -\infty. 
\end{equation}
We have chosen to illustrate two specific sub--cases, namely, $\alpha = 0.7$ and $\alpha = 0.9$. 
These correspond to the left--hand and right--hand plots of Figures~\ref{F:3} and ~\ref{F:3b} respectively. We observe that large stability regions exist,  except in the limit $x\rightarrow1$ (with $y\neq1$). 
It appears that the difference between $\alpha=0.7$ and $\alpha=0.9$ is qualitatively negligible. However, of particular interest is the region \emph{very close} to the asymptote at $x=1$; where we again have a `pit'. This leads to a preferred choice of the parameters $a, m_{\pm}$ which in turn leads to regions of maximal stability.

Note that in this situation
\begin{eqnarray}\label{stabilityineq00}
    \frac{m_{s}(R_{0})}{R_{0}} 
    &=& -\sqrt{1-\frac{2\alpha m_{*}\e^{-a/R_{0}}}{R_{0}}} 
           - \sqrt{1-\frac{2m_{*}\e^{-a/R_{0}}}{R_{0}}}
    \nonumber\\[10pt]
    &=& -\sqrt{1-\alpha x} -\sqrt{1-x} = -\sqrt{1-\alpha} + \mathcal{O}(1-x). 
    \end{eqnarray}
    
 Thus for $\alpha<1$ the energy condition violations are minimized (though no longer arbitrarily small) as the wormhole throat approaches the location of what would be a horizon in the bulk spacetime~\cite{small1,small2}.

\subsection{Specific asymmetry: $a_{+}\neq a_{-}$ while $m_{+}=m_{-}$}

Let us now suppose $a_{+}\neq a_{-}$ and $m_{+}=m_{-}=m$. Hence we are now performing surgery between two asymptotically Minkowski black holes with identical masses but different exponential suppression parameters. To develop a  tractable analysis, define
\begin{equation}
a_*=\min\{a_+,a_-\}; \qquad\qquad
 \beta = {\max\{a_+,a_-\}\over \min\{a_+,a_-\} } \geq 1.
\end{equation}

We then have:
\begin{eqnarray}\label{asymm2}
    R_{0}^{2}\left[\frac{m_{s}(R_{0})}{R_{0}}\right]^{''} &\geq& F_{4}(R_{0}, m, a_{-},\beta) \nonumber \\
    && \nonumber \\
    &=& \frac{\left[m\,\e^{-\frac{\beta a_*}{R_{0}}}\left(1-\frac{\beta a_{*}}{R_{0}}\right)\right]^{2}}{R_{0}^{2}\left[1-\frac{2m\,\e^{-\frac{\beta a_{*}}{R_{0}}}}{R_{0}}\right]^{\frac{3}{2}}} 
    - \frac{m\beta a_{*}\e^{-\frac{\beta a_{*}}{R_{0}}}\left(4-\frac{\beta a_{*}}{R_{0}}\right)}{R_{0}^{2}\sqrt{1-\frac{2m\,\e^{-\frac{\beta a_{*}}{R_{0}}}}{R_{0}}}} \nonumber \\
    && \nonumber \\
    && + \frac{\left[m\,\e^{-\frac{a_{*}}{R_{0}}}\left(1-\frac{a_{*}}{R_{0}}\right)\right]^{2}}{R_{0}^{2}\left[1-\frac{2m\,\e^{-\frac{a_{*}}{R_{0}}}}{R_{0}}\right]^{\frac{3}{2}}} - \frac{ma_{*}\e^{-\frac{a_{*}}{R_{0}}}\left(4-\frac{a_{*}}{R_{0}}\right)}{R_{0}^{2}\sqrt{1-\frac{2m\,\e^{-\frac{a_{*}}{R_{0}}}}{R_{0}}}} \ .
\end{eqnarray}
\enlargethispage{40pt}

The stability analysis may now be simplified by employing the two \emph{dimensionless} parameters $x=\frac{2m}{R_0} \,\e^{-a_*/R_0}$ and $y=\frac{a_{*}}{R_0}$ to re--express this stability condition
as a function of these dimensionless parameters. Specifically
\begin{equation}
\label{asymm3}
    F_{4}(x,y) = 
    \frac{\left[x\e^{(1-\beta)y}\left(1-\beta y\right)\right]^{2}}
    {4\left(1-x \e^{(1-\beta)y}\right)^{\frac{3}{2}}} 
    - \frac{\beta xy\,\e^{(1-\beta) y}(4-\beta y)}{2\sqrt{1-x\e^{(1-\beta) y}}} 
    + \frac{\left[x(1-y)\right]^{2}}{4\left(1-x\right)^{\frac{3}{2}}} 
    - \frac{xy(4-y)}{\sqrt{1-x}} \ .
\end{equation}
Notice that  the square root in the denominator implies $0<x<1$. We may however once again assert $0<y<1$ if the bulk spacetimes contain horizons, while $1<y<\infty$ is permitted if the bulk spacetimes are horizon-free.

Since by construction $\beta\geq 1$ it is easy to check that
\begin{equation}
\lim_{x\to1} F_{4}(x, y\neq 1) = +\infty; \qquad\qquad
\lim_{x\to1} F_{4}(x, y=1) = -\infty. 
\end{equation}

\begin{figure}[htb!]
\begin{subfigure}{.5\textwidth}
\includegraphics[scale=0.450]{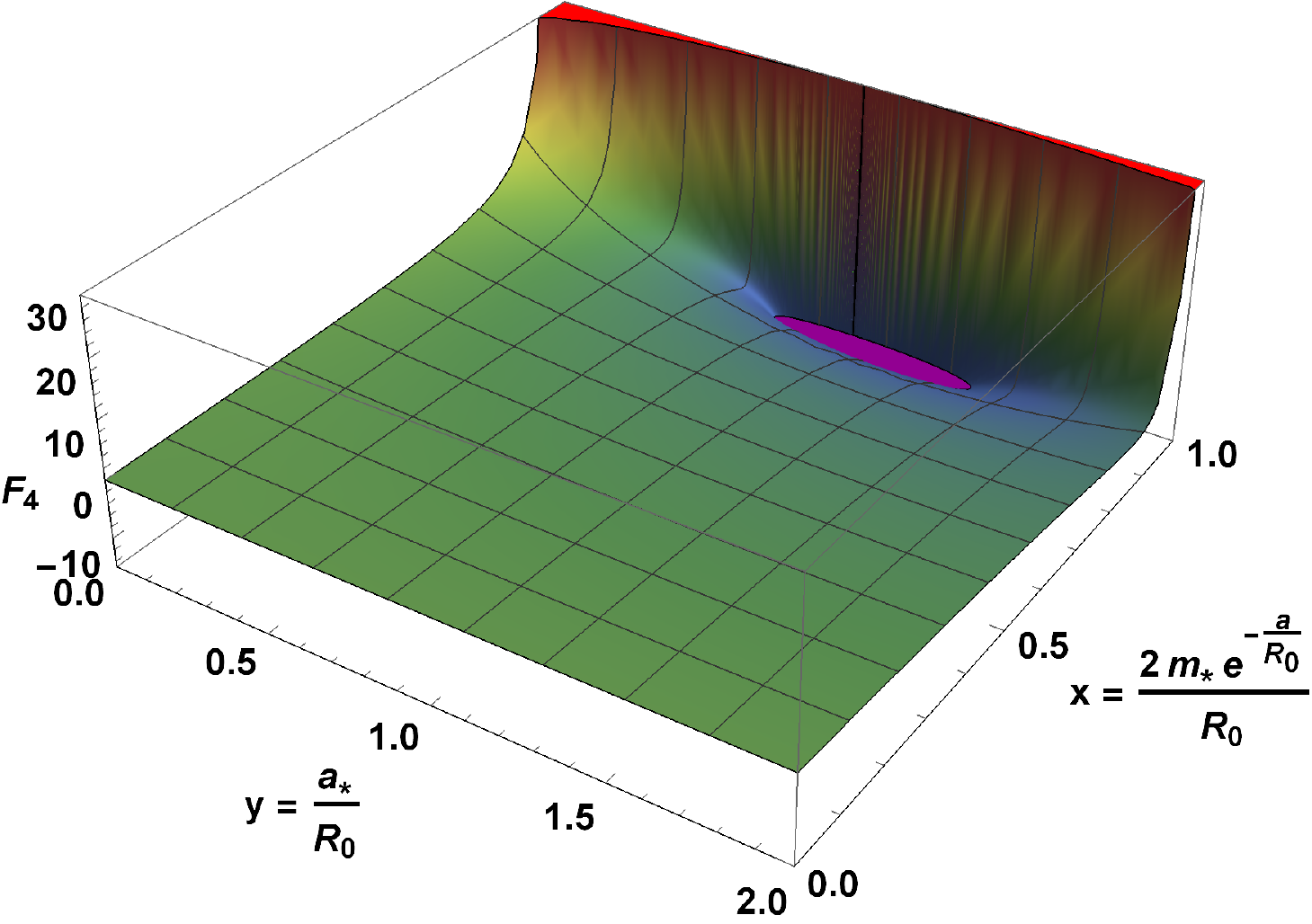}
\caption{$\beta = 1.2$}
\end{subfigure}
\begin{subfigure}{.5\textwidth}
\includegraphics[scale=0.450]{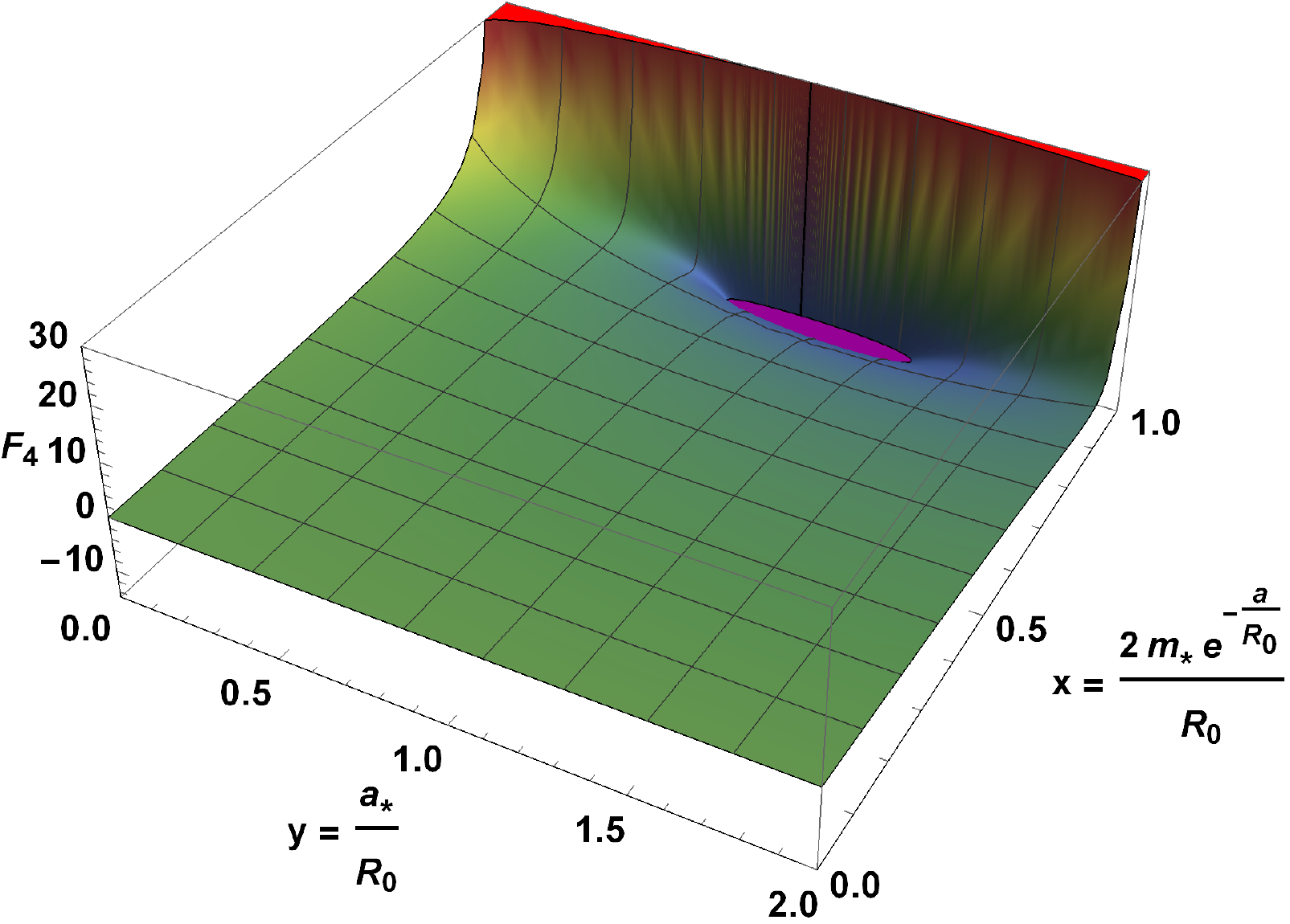}
\caption{$\beta=1.4$}
\end{subfigure}
\caption{
Stability analysis for the asymmetry $\max\{a_{+},a_-\} =\beta \min\{a_+,a_-\} = \beta a_{*}$, with $\beta >1$, $m_{+} = m_{-}=m$. The stability region lies above the surface $F_{4}(x, y)$. The red and purple regions indicate where the function $F_4(x,y)$ departs the range $(-20,+30)$.}
\label{F:5a}
\end{figure}

\begin{figure}[htb!]
\begin{subfigure}{.5\textwidth}
\includegraphics[scale=0.215]{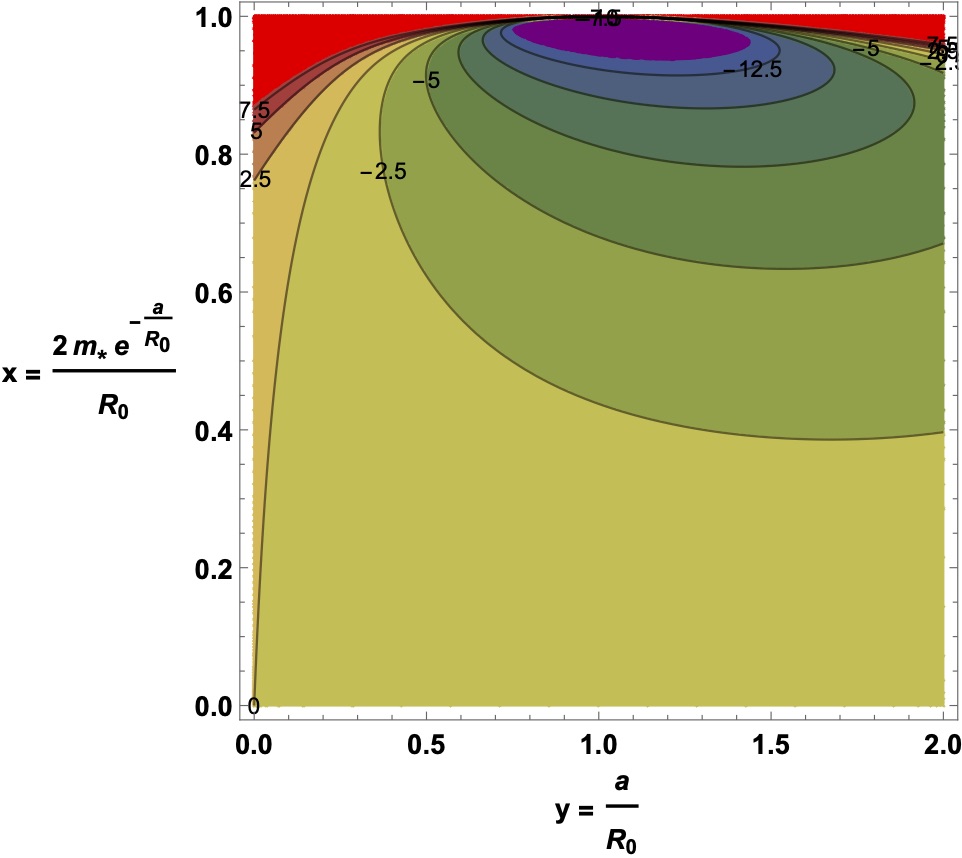}
\caption{$\beta = 1.2$}
\end{subfigure}
\begin{subfigure}{.5\textwidth}
\includegraphics[scale=0.215]{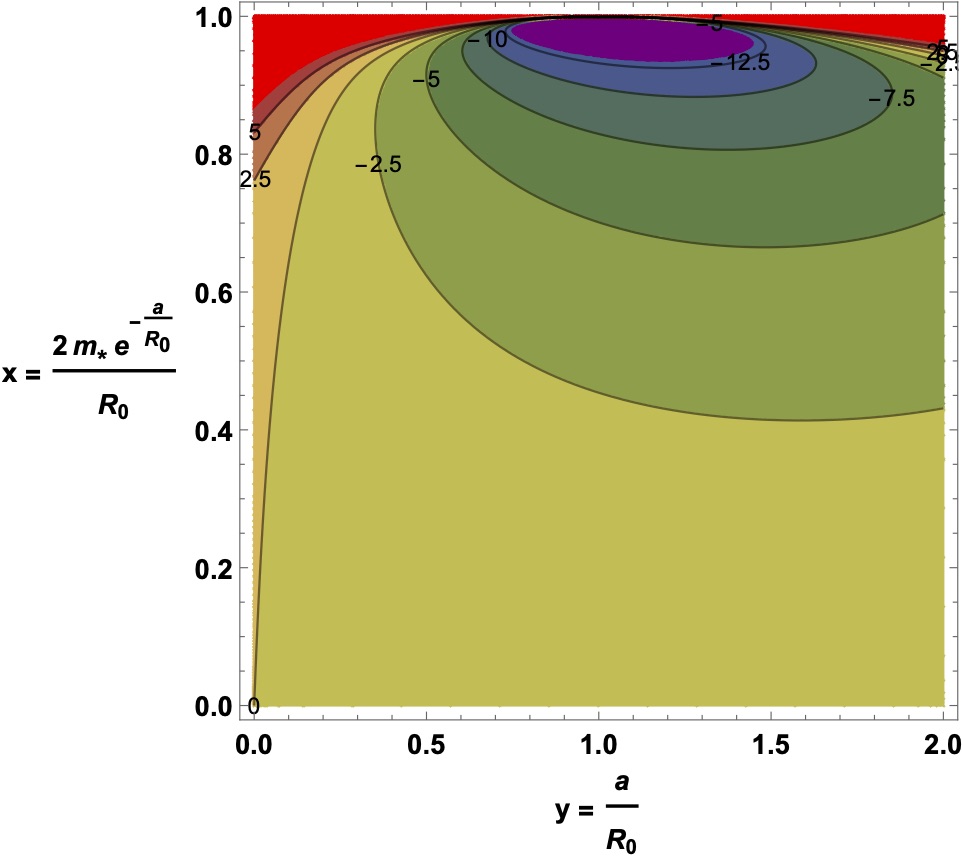}
\caption{$\beta=1.4$}
\end{subfigure}
\caption{
Contour plots:
Stability analysis for the specific asymmetry $\max\{a_{+},a_-\} =\beta \min\{a_+,a_-\} = \beta a_{*}$, with $\beta >1$, $m_{+} = m_{-}=m$. These are contour plots for the function $F_{4}(x, y)$. The purple region indicates the `pit' where the function 
function $F_4(x,y)$ is strongly negative.
The red region indicates the region of decreased stability  where the function 
function $F_4(x,y)$ is strongly positive.}
\label{F:5b}
\end{figure}

For illustrative purposes we present the specific cases $\beta=1.2$ and $\beta=1.4$. These correspond to the left--hand and right--hand plots of Figure~\ref{F:5a}  and Figure~\ref{F:5b}  respectively. We have large stability regions other than in the  limit $x\to1$ (with $y\neq1$). There is again a `pit' in the vicinity of $(x,y)\approx(1,1)$. 

We note that in this situation
\begin{eqnarray}\label{stabilityineq0xxx}
    \frac{m_{s}(R_{0})}{R_{0}} 
    &=& 
    -\sqrt{1-\frac{2m\,\e^{-\beta a_{*}/R_{0} }}{R_{0}}} 
   - \sqrt{1-\frac{2m\,\e^{-a_{*}/R_{0}}}{R_{0}}} \,,
   \nonumber\\[10pt]
   &=& - \sqrt{ 1 - \e^{(1-\beta) y} x} - \sqrt{1-x} 
   \nonumber\\[5pt]
   &=&  - \sqrt{ 1 - \e^{(1-\beta) y}} + \mathcal{O}(1-x).
    \end{eqnarray}
    Thus for $\beta>1$ the energy condition violations are minimized (though no longer arbitrarily small) as the wormhole throat approaches the location of what would be a horizon in the bulk spacetime~\cite{small1,small2}.

\section{Discussion and Conclusions}

In this work we have used a novel regular black hole model based on exponential mass suppression to construct a thin-shell wormhole using the cut-and-paste technique.
The construction under consideration provides an example of a black hole mimicker.
(The smaller the value of the mass suppression parameter $a$ and the closer the location of the wormhole throat to the Schwarzschild radius, the better this this model is to mimicking a standard Schwarzschild black hole.) 
For suitable choices of parameters, the wormhole under consideration was found to violate the null energy condition in the bulk spacetime, whereas the strong energy condition is satisfied in this region.
The wormhole construction was analysed via the thin-shell formalism, allowing the four-velocity of the wormhole throat to be calculated along with the junction surface unit normal vectors, the extrinsic curvature, and the junction surface stress-energy.
The surface energy at the wormhole junction throat was found to be negative, and so, much like other traversable wormholes, exotic matter would be needed to keep the wormhole throat open.
\enlargethispage{40pt}

We found that this class of wormholes permits a clean and quite general stability analysis, with wide swathes of stable behaviour. Furthermore the stability plateau exhibits a `pit' of enhanced stability when the wormhole throat is close to where a near-extremal horizon would have existed in the bulk spacetime before applying `cut-and-paste' surgery. 
Finally we found that the quantity of exotic matter needed to support the wormhole throat could be minimized (and in some cases made arbitrarily small) by suitable choice of parameters.

\section*{Acknowledgements}

TB was supported by a Victoria University of Wellington MSc scholarship, 
and was also indirectly supported by the Marsden Fund, 
via a grant administered by the Royal Society of New Zealand.
\\
FSNL acknowledges support from the Funda\c{c}\~{a}o para a Ci\^{e}ncia e a Tecnologia (FCT) Scientific Employment Stimulus contract with reference CEECIND/04057/2017, and research grants No. UID/FIS/04434/2020, No. PTDC/FIS-OUT/29048/2017 and No. CERN/FIS-PAR/0037/2019.
\\
AS was supported by a  Victoria University of Wellington PhD scholarship,
and was also indirectly supported by the Marsden Fund, 
via a grant administered by the Royal Society of New Zealand.
\\
MV was directly supported by the Marsden Fund, via a grant administered by the Royal Society of New Zealand.

\bigskip

\end{document}